\documentclass[11pt]{article}
\pdfoutput=1
\usepackage{verbatim}
\usepackage{amssymb}
\usepackage{xcolor}
\usepackage[centertags]{amsmath}
\usepackage[square, comma, sort&compress,numbers]{natbib}	
\usepackage{array,multirow}
\usepackage{graphicx}
\usepackage{bbold}
\usepackage{braket}
\usepackage{wrapfig}
\usepackage{float}
\usepackage{soul}
\usepackage{url}
\numberwithin{equation}{section}
\usepackage{tikz,pgf}
\usetikzlibrary{shapes}
\usetikzlibrary{calc}
\usetikzlibrary{decorations.pathmorphing}
\usetikzlibrary{decorations.pathreplacing,shapes.misc}
\usetikzlibrary{positioning}
\usetikzlibrary{arrows}
\usetikzlibrary{decorations.markings}
\usetikzlibrary{shadings}
\usetikzlibrary{intersections}
\usepackage{jheppub}
\numberwithin{equation}{section} \makeatletter
\@addtoreset{equation}{section}

\newcommand{\be}{\begin{equation}}
\newcommand{\ee}{\end{equation}}
\newcommand{\deffl}{\mathrel{\mathop:}=}
\newcommand{\dps}{\displaystyle}
\newcommand{\bL}{\mathbf{L}}
\newcommand{\bW}{\mathbf{W}}
\newcommand{\RR}{\mathbb{R}}
\newcommand{\ZZ}{\mathbb{Z}} 
\newcommand{\varC}{\mathcal{C}}
\newcommand{\varM}{\mathcal{M}}
\newcommand{\varW}{\mathcal{W}}
\newcommand{\varR}{\mathcal{R}} 
\newcommand{\Id}{\mathbf{1}}
\newcommand{\mf}[1]{\mathfrak{#1}}
\renewcommand{\sl}{\mathfrak{sl}} 
\renewcommand{\a}{\Omega}

\DeclareMathOperator{\AdS}{AdS}
\DeclareMathOperator{\Tr}{Tr}

\DeclareMathOperator{\SL}{SL}
\DeclareMathOperator{\SU}{SU}
\DeclareMathOperator{\inst}{inst}

\makeatletter
\def\@fpheader{\vspace{-.1cm}}
\makeatother

\title{Wilson lines construction of $\mathfrak{sl}_3$ toroidal conformal blocks}

\author{Vladimir Belavin, Pietro Oreglia, J. Ramos Cabezas}

\affiliation{Physics Department, Ariel University, Ariel 40700, Israel.}

\emailAdd{vladimirbe@ariel.ac.il, pietro.oreglia@msmail.ariel.ac.il, juanra@ariel.ac.il}

\abstract{
We study $\varW_3$ toroidal conformal blocks for degenerate primary fields in AdS/CFT context. 
In the large central charge limit $\varW_3$ algebra reduces to $\mathfrak{sl}_3$ algebra and $\mathfrak{sl}_3$ blocks are defined as contributions to $\varW_3$ blocks coming from the generators of $\mathfrak{sl}_3$ subalgebra. 
We consider the construction of $\mathfrak{sl}_3$ 
toroidal blocks in terms of Wilson lines operators of $3d$ Chern-Simons gravity in the thermal AdS$_3$ space-time. According to the correspondence, 
degenerate primary fields are associated with Wilson lines operators acting in the corresponding finite-dimensional $\mathfrak{sl}_3$ representations. We verify 
this dual construction 
for one-point toroidal block using
 $\mathfrak{sl}_3$ tensor  technique in the bulk theory and an algorithm based on AGT correspondence in the boundary CFT. 
} 

\begin{document}
\maketitle
\flushbottom
\section{Introduction} 

Conformal blocks (CB's) determine the holomorphic contributions to the correlation functions that appear after fixing the OPE channels~\cite{Belavin:1984vu}.
CB's are generally fixed by the symmetry algebra and depend on the topology of the CFT Riemann surface.
In this work, we are interested in $\varW_3$ one-point conformal block  on the torus in the large central charge limit. 

The one-point correlation function of a primary field $\Phi_{\alpha_1}$ with conformal dimension $h_1$ on the torus is defined as\footnote{Throughout this paper we omit factor $(q\bar{q})^{-\frac{c}{24}}$ which can be easily restored.} 
\begin{equation} 	
\braket{\Phi_{\alpha_1}(z_1, \bar{z}_1)}= \Tr_{\alpha}\Big( q^{\bL_0} \bar{q}^{\bar{\bL}_0}\Phi_{\alpha_1}(z_1, \bar{z}_1)    \Big)=\sum_{\alpha} C_{\alpha \alpha_1 \alpha} |\mathcal{F}(\alpha, \alpha_1, q)|^2 \; ,
\label{ni2}	 
\end{equation}
where $\Tr_{\alpha}$ is the trace taken over a module of the symmetry  algebra associated with the primary field $\Phi_{\alpha}$ in the intermediate OPE channel,
$q$ is the elliptic parameter of the torus $q=e^{2\pi i \tau }$ and $\bL_0$ is the generator of the algebra satisfying $\bL_0\ket{h_1} =h_1\ket{h_1}$. Here $\mathcal{F}(\alpha, \alpha_1, q)$ is the one-point holomorphic toroidal conformal block (for more details, see, e.g., \cite{He:2012bi, Hadasz:2009db, DiFrancesco:1997nk}).

In recent years, the AdS$_3$/CFT$_2$ provided a new formulation of
conformal blocks in terms of geodesic networks in $\AdS$ space-time, see e.g., \cite{Hartman:2013mia, Fitzpatrick:2014vua, Caputa:2014eta, Hijano:2015rla, Fitzpatrick:2015zha, Alkalaev:2015wia, Hijano:2015qja, Hijano:2015zsa,  Alkalaev:2015lca, Alkalaev:2015fbw, Banerjee:2016qca, Gobeil:2018fzy, Hung:2018mcn, Alekseev:2019gkl, RamosCabezas:2020mew}, or \emph{Wilson network operators} of Chern-Simons gravity, which will be the subject  of the present consideration.\footnote{For recent development on CB's in the holography context, see, e.g.,
~\cite{ deBoer:2013vca, Ammon:2013hba, deBoer:2014sna, Hegde:2015dqh, Melnikov:2016eun, Bhatta:2016hpz, Besken:2017fsj, Hikida:2017ehf, Hikida:2018eih, Hikida:2018dxe, Besken:2018zro, Bhatta:2018gjb, DHoker:2019clx, Castro:2018srf, Kraus:2018zrn, Hulik:2018dpl, Castro:2020smu, Chen:2020nlj,Belavin:2022bib}.}

The latter approach is based on the higher-spin version of AdS/CFT correspondence~\cite{Gaberdiel:2010pz}, which, in general, identifies the minimal model cosets 
\be
\frac{\SU(N)_k \oplus \SU(N)_1}{SU(N)_{k+1} }
\label{eqn:intro1} 
\ee
in \emph{'t Hooft limit}, as the holographic dual of the higher spin theory~\cite{Prokushkin:1998bq}. 
Here we study a different limit, the so-called 
\emph{semiclassical limit}, where we have, from the bulk side, an $\SL(N) \times \SL(N)$ Chern-Simons theory, while from the CFT side, a non-unitary $\varW_N$ CFT model considered in the large central charge limit, with $N$ fixed.
We recall  that primary fields in the $\varW_N$ model are labelled by a pair of $\SL(N)$ highest-weights $(\Lambda_+, \Lambda_-)$, which are both highest-weights of a finite-dimensional representation of $\SL(N)$~\cite{Perlmutter:2012ds}. One can identify two different kinds of primaries fields:
\begin{itemize}
\item \emph{heavy operators}, identified by $(0, \Lambda_-)$, which have scale dimensions $\Delta \sim c$, and correspond to flat $\SL(N) \times \SL(N)$ connections in the bulk;
\item \emph{light operators}, labelled by $(\Lambda_+, 0)$, whose scale dimensions go as $\Delta \sim \mathit{o}(1)$, and are related to perturbative matter in the bulk. 
\end{itemize}
In this paper we focus on the light operators. This allows us to consider the large $c$ limit by restricting the set of generators, which in general can be written as $\{L_{n},\,  W^3_{n},\, \dots, \, W^{N}_{n}\}$, to those of the types
\be
L_{n} \; \text{ for }\;  |n| < 2 \; , \qquad W^s_{n} \; \text{ for }\; |n| < s \; , 
\ee 
with $s = 3, \dots, N$. By looking at the commutation relations (for explicit form in $N=3$ case, see sec.~\ref{prel}), it can be seen that these operators can be identified as the generators of the $\sl_N$ algebra. In the spherical topology (for $N=2,3$) this program has been implemented in~\cite{Besken:2016ooo}. Regarding the toroidal topology, it was shown in \cite{Alkalaev:2020yvq} that $\mathfrak{sl}_2$ one-point toroidal block $\mathcal{F}(-j, -j_1,q)_{\sl_2}$, where $-j$ and $-j_1$ are respectively the intermediate and external conformal dimensions, is given by 
\begin{equation} 
 \mathcal{F}(-j, -j_1,q)_{\sl_2}=\Tr_{j}\Big{(}W_{j    }[z_b, z_b+2\pi \tau]I_{j;j, j_1} \Big{)}\otimes W_{j_1}[z_b, z_1]\ket{lw}_1 \; .
\label{prop1} 
\end{equation}
 In the dual description $j$ and $j_1$ are the spins of $\mf{sl}_2$ Chern-Simons gauge group. The trace $\Tr_{j}(\cdots)$ is taken over the representation with spin $j$, $z_1$ is the point on the boundary of the solid torus, which is a geometric representation of the thermal AdS$_3$, $z_b$ is an arbitrary point in the bulk of AdS$_{3}$, $\ket{lw}_1$ is the lowest-weight state of the representation with spin $j_1$ and $I_{j;j,j_1}$ is the intertwining operator associated with the representations $j$ and $j_1$. The factors $W_i[x,y]$ (for $i=j,j_1$; $x=z_b$ and $y= z_b+2\pi \tau,z_1 $) denote the Wilson line operators
\begin{equation}
W_{i}[x,y]=\mathcal{P}\exp \bigg(-\int_{x}^{y} \a \bigg) = \exp \left( (x-y) (L_1+\frac{1}{4} L_{-1}  ) \right) \; .
\label{prop2} 
\end{equation}
Here $L_{1}$ and $L_{-1}$ are the lowering and raising operators of $\sl_2$ algebra in the representation with spin $i$. 

In this work we consider the $\varW_3$ algebra; 
we are interested in fully degenerate primary fields $\Phi_{\alpha}(z, \bar{z})$, characterized by a pair of quantum numbers $(h_{\alpha}, q_{\alpha})$ (\emph{conformal dimension} and $\varW_3$ \emph{charge}, respectively).\footnote{ 
Explicit form of  $(h_{\alpha}, q_{\alpha})$ in terms of $\alpha$ will be given below.} 
In the semiclassical limit $c \rightarrow \infty$, or $b\rightarrow 0$ in Toda-like parameterization~\cite{Fateev:2007ab}, $c= 2 +24 (b +1/b)^2$,  the parameter $\alpha$ corresponding to the intermediate primary field $\Phi_{\alpha}$ in~\eqref{ni2} is given by
\begin{equation}
\alpha \rightarrow -bj\,, \quad  j=m_1w_1 +m_2w_2 \; ,
\label{alpha2.0} 
\end{equation}
where $w_1$, $w_2$ are $\sl_3$ fundamental weights and $m_1$, $m_2$ are non-negative integers.
For the external primary $\Phi_{\alpha_1}$ characterized by $(h_{\alpha_1}, q_{\alpha_1})$
\begin{equation}
\alpha_1 \rightarrow -bj_1\,, \quad  j_1=aw_1 \; ,  
\label{alpha3} 
\end{equation}
where $a$ is a non-negative integer. The external field $\Phi_{\alpha_1}$ is restricted to the so-called semi-degenerate form, to avoid the multiplicities problem of conformal blocks in $\varW_3$ CFT (for details\footnote{We will also comment on this problem in our conclusions in section~\ref{conclusion}, as a generalization of the present work.} and consequences of this problem see, e.g.,~\cite{Belavin:2016qaa,  Fateev:2007ab}). 
In what follows, we focus on the one-point blocks with external fields satisfying~\eqref{alpha3}, which can be equivalently written as
\begin{equation}
\bW_{-1}\Phi_{\alpha_1} (0,0)\ket{0}= \frac{3q_1}{2h_1}\bL_{-1}\Phi_{\alpha_1}(0,0)\ket{0}\;.
\label{alpha4}
\end{equation}

The outline of the paper is the following. In section~\ref{prel}, we recall the necessary facts about $\varW_3$ CFT and $\AdS_3$ Chern-Simons gravity. 
In sections~\ref{1ptw3} and~\ref{agt1}, we compute $\mathcal{F}(\alpha, \alpha_1,q)_{\sl_3}$ and check it by comparing with the large central charge limit of $\varW_3$ block of light operators, using the algorithm based on AGT relation. To this end we propose a relation between the light $\varW_3$ and  $\mathfrak{sl}_3$ one-point blocks, which is similar to the one existing between $\mathfrak{sl}_2$ and Virasoro light blocks in the large central charge limit~\cite{Alkalaev:2016fok}. In section~\ref{wilsonline}, we describe the dual construction for $\mathfrak{sl}_3$ one-point block in terms of the Wilson lines operators. We find that the expression obtained in section~\ref{1ptw3} can be represented by the lhs of~\eqref{prop1}. Our conclusions are collected in section~\ref{conclusion}. Appendices \ref{apA}, \ref{apweylchafor}, \ref{apagtrelation}, \ref{wilexamples} contain some technical details related to $\mathfrak{sl}_3$ matrix elements, $\mathfrak{sl}_3$ Weyl character formula, AGT relation, and Wilson lines description of conformal blocks respectively.

\section{Preliminaries: $\AdS_3$ and $\varW_3$ CFT} 
\label{prel} 
\subsection{$\varW_3$ Conformal Field Theory} 
\label{theory1} 
The symmetry of the $\varW_3$ CFT is generated by the energy-momentum tensor $\mathbf{T}(z)$ (a spin-$2$ current) and the additional spin-3 current $\mathbf{W}(z)$. Their expansions as Laurent series read
\begin{equation}
\mathbf{T}(z)= \sum_{n=-\infty}^{\infty} \frac{\bL_n}{z^{n+2}},\qquad \bW(z)= \sum_{n=-\infty}^{\infty} \frac{\bW_n}{z^{n+3}}.
\end{equation}
The modes $\bL_n$ and $\bW_m$ generate the $\varW_3$ algebra, that is, they satisfy the commutation relations 
\begin{equation}
 \begin{split}
\left[\bL_n, \bL_m\right]		& =(n-m) \bL_{n+m}+\dps \frac{c}{12}(n^3-n) \delta_{n+m,0} \; , \\ 
\left[\bL_n, \bW_m\right]		& =(2n-m) \bW_{n+m} \; , \\ 
\left[\bW_n, \bW_m\right]		& = \frac{c}{3\cdot5!}(n^2-1)(n^2-4)n \delta_{n+m,0}+\frac{16}{22+5c}(n-m) \mathbf{\Lambda}_{n+m} + \\ 
						& \qquad + \frac{(n-m)}{30}\left( 2m^2 +2n^2 - m n -8  \right)\bL_{n+m} \; ,
\end{split}
\end{equation}
where
\begin{equation}
\mathbf{\Lambda}_{m} = \sum_{p \leq - 2} \bL_{p} \bL_{m-p} + \sum_{p \geq - 1} \bL_{m-p} \bL_{p} - \frac{3(m+2)(m+3)}{10} \bL_{m} \; .
\label{nl} 
\end{equation}
In the limit $c\rightarrow \infty$ these commutation relations reduce to the ones of the $\sl_3$ algebra, generated by
\begin{equation} 
\{ \bL_{-1}, \bL_0, \bL_1, \bW_{-1}, \bW_1, \bW_0, \bW_{-2}, \bW_2 \} \; ,
\label{sl3generators} 
\end{equation}
that satisfy
\begin{equation}
\begin{split}
[\bL_n, \bL_m] 		& = (n-m) \bL_{n+m} \; , \\
[\bL_n, \bW_m] 	& = (2n-m)\bW_{n+m} \; ,  \\ 
[\bW_n, \bW_m] 	& = (n-m)\Big(\frac{1}{15}(n+m+2)(n+m+3)-\frac{1}{6}(n+2)(m+2)\Big) \bL_{n+m} \; . 
\end{split}
\label{eqn:W3.1} 
\end{equation}
\paragraph{$\varW_3$ primary fields.} 
The conformal dimension $h_\alpha$ and charge $q_\alpha$ of a primary field $\Phi_{\alpha}$ in a $\varW_3$ CFT are expressed as
\be
h_{\alpha} = \frac{1}{2}(\alpha, \, 2Q-\alpha) \, , \qquad q_{\alpha} = i\sqrt{\frac{48}{22 + 5c}}\prod_{i=1}^{3}(e_i, \, \alpha - Q) \; , 
\label{dim} 
\ee
Here 
$Q = (b + \frac{1}{b})(w_1 + w_2)$, $e_i$ are the weights of the fundamental representation
\begin{equation}  
e_1= w_1\, , \quad  e_2= w_2-w_1\, , \quad e_3= -w_2 \; ,
\label{voffundamentalr}
\end{equation}
and $\alpha$ is a vector on the root space given by
\begin{equation} 
\alpha_{r_1r_2 s_1 s_2} = b  \big( (1-r_1)w_1 + (1-r_2)w_2 \big)+ \frac{1}{b} \big( (1-s_1)w_1 + (1-s_2)w_2 \big) \; ,
\label{fulldeal} 
\end{equation}
with $r_1$, $r_2$, $s_1$ and $s_2$ positive integers. In the large central charge limit, we have that~\eqref{fulldeal} becomes
\begin{equation}
\alpha \rightarrow -bj, \quad  j=m_1w_1 +m_2w_2 \; ,
\label{alpha2} 
\end{equation}
for which the conformal dimension and the charge assume the values
\begin{equation}
h_{\alpha}= -m_1-m_2, \qquad q_{\alpha} = \frac{i}{3}\sqrt{\frac{2}{5}}(m_2-m_1) \; .
\label{dimd} 
\end{equation}

\paragraph{$\varW_3$ module.}

A $\varW_3$ \emph{highest-weight vector} $\ket{h_{\alpha}, q_{\alpha}}$ given by
\begin{equation}
\ket{h_{\alpha}, q_{\alpha}} = \lim_{z \rightarrow 0}  \Phi_{\alpha}(z) \ket{0}\; , 
\end{equation}
satisfies the conditions
\begin{gather}
\bL_0 \ket{h_{\alpha}, q_{\alpha}} = h_{\alpha} \ket{h_{\alpha}, q_{\alpha}} \; , \qquad \bW_0 \ket{h_{\alpha}, q_{\alpha}} = q_{\alpha} \ket{h_{\alpha}, q_{\alpha}} \; , \\
\bL_{n}\ket{h_{\alpha}, q_{\alpha}} = \bW_{n}\ket{h_{\alpha}, q_{\alpha}} = 0 \; , \qquad n > 0 \; .
\end{gather}
The $\varW_3$ \emph{module} associated with this highest-weight vector is spanned by the basis of descendant states
\begin{equation} 
\mathcal{L}_{-I}   \ket{h_{\alpha}, q_{\alpha}}  =  \bL_{-i_1} \dots \bL_{-i_m} 
   \bW_{-j_1} \dots \bW_{-j_n} \ket{h_{\alpha}, q_{\alpha}} \, , \quad I= \{i_1, \dots, i_m; j_1,\dots, j_n \} \; , \\
\label{w3module} 
\end{equation}
with 
\be
1 \le i_1 \le \dots \le i_m \, , \quad  1 \le j_1 \le \dots \le j_n \; .
\ee
The sum
\begin{equation}
 N= \sum_{i_a, j_b \in I} i_a + j_b  
\end{equation}
is called \textit{level} of the state. Similarly, the $\mathfrak{sl}_3$ \emph{module} is defined when in the set $I$, the indices are restricted to
\begin{equation} 
i_1,\dots, i_m=1 \;, \quad  j_1, \dots , j_n=1,2 \; ,
\label{sl3module} 
\end{equation}
that is, when descendants are generated only by $\bL_{-1}$, $\bW_{-1}$ and $\bW_{-2}$. For the basis states of $\mathcal{W}_3$ and $\mathfrak{sl}_3$ modules, we introduce the following convenient notation 
\begin{equation} 
\mathcal{L}_{-I}  \ket{h_{\alpha}, q_{\alpha}} =  \ket{h_{\alpha}, q_{\alpha} , N} \; .
\label{descendentnot} 
\end{equation}

\subsection{Brief review of 3d Chern-Simons gravity theory} 
It is known~\cite{witten19882+} that the $2+1$-dimensional Einstein-Hilbert action $S$ with a negative cosmological constant can be written in terms of the Chern-Simons action
\be
S = S_{\footnotesize \text{CS}}[A] - S_{\footnotesize \text{CS}}[\bar{A}] \; , 
\ee
where
\be
S_{\footnotesize \text{CS}}[A] = \frac{k}{4\pi} \int_{\varM^3} \Tr \Big( A \wedge d A + \frac{2}{3}A \wedge A \wedge A\Big) \; . 
\label{eqn:CS1}	
\ee
The constant $k$ is related to the $3$-dimensional Newton constant $G_3$ as 
\be
k = \frac{r}{4 G_3} \; , 
\ee
with $r$ the $\AdS_3$ radius, $\Tr$ stands for the invariant Killing form, $\varM^3$ is the $3$-dimensional space-time and $A$ ($\bar{A}$) is the (anti-) chiral gauge  \textit{connection}, that is, a one-form valued in the gauge group, given by a composition of vielbein and spin-connection. In the case we are interested in, that is the $\SL(N)\times \SL(N)$ Chern-Simons theory, $A$ is an $\SL(N)$ connection. The equations of motion that follow from~\eqref{eqn:CS1} are given by (the same equations are  valid for the anti-chiral connection)
\be
dA + A \wedge A = 0 \; , 
\label{eqn:CS2}	
\ee
which are the flatness conditions on the connection. It can be shown~\cite{Castro:2011iw} that, by choosing the proper boundary conditions, in the local coordinates $x^{\mu}= (\rho, z, \bar{z})$ with $\rho \ge 0$ the radial one, the solution of~\eqref{eqn:CS2} can be written as the gauge-transformed $\Omega$:
\be 
 A = U^{-1} \Omega U + U^{-1}dU \; , 
 \ee
 with 
 \be 
 \Omega = \Big(L_1 - 2 \pi \frac{6 T(z)}{c}L_{-1}\Big) d z \; , \qquad U = e^{\rho L_0} \; . 
 \ee
Here $T(z)$ is the holomorphic boundary energy-momentum tensor, while $c$ is the central charge, defined through the Brown-Henneaux relation $c= \frac{3r}{2G_3}$~\cite{Brown:1986nw}. $L_1$, $L_{-1}$ are two of the generators of the $\sl_2$ subalgebra of the gauge algebra  $\sl_N$. We are interested in the space-time with periodic time conditions, that is to say, the thermal $\AdS_3$, in which case the stress-energy tensor is 
 \be
 T(z) = -\frac{c}{48 \pi} \; , 
 \ee
 and the connection $\Omega$ becomes 
 \be
 \Omega = \Big( L_1 + \frac{1}{4}L_{-1}\Big) dz \; , 
 \ee
 together with the identifications $z \sim z + 2\pi$ and $z \sim z + 2\pi \tau$, with $i\tau \in \RR_{<0}$. 
 
Finally, we define the \textit{Wilson line operators}. Given a chiral connection $A$ defined in a representation $R$ of $\mathfrak{sl}_N$ and a path $l$ that connects two points $x_1, x_2 \in \varM^3$, we define the Wilson line operator $W_R[l]$ as 
\begin{equation}
W_R[l]= \mathcal{P} \exp \Big[ \int_l A\Big] \; ,
\label{wlop}
\end{equation}
where $\mathcal{P}$ denotes the path-ordering operator. 

\section{One-point conformal block from CFT} 
\label{1ptw3} 
In this section, we define the holomorphic $\mathcal{W}_3$ and $\mathfrak{sl}_3$ one-point conformal blocks\footnote{For further discussions related to the $\mathcal{W}_3$ one-point conformal blocks see also \cite{Chang:1991ht, He:2012bi, Hadasz:2009db}.}. 
The result of the calculation of 
the $\mathcal{W}_3$ block will be given in the next section. 
The holomorphic $\mathcal{W}_3$ one-point conformal block $\mathcal{F}(\alpha, \alpha_1, q)$
is given by
\begin{equation} 
\mathcal{F}(\alpha, \alpha_1, q)  = \frac{1}{ \bra{h_0,q_0} \Phi_{\alpha_1}(z) \ket{h_0,q_0}} \sum_{\substack{M,N=0 \\ M=N}}\bra{h_0, q_0, M} q^{L_0} \Phi_{  \alpha_1}(z) \ket{h_0,q_0, N} G_{MN}^{-1} \; , 
\label{1pt} 
\end{equation}
where $ h_{\alpha}:=h_0$, $ q_{\alpha}:=q_0$, the states $\ket{h_0,q_0, N} $ belong to the $\mathcal{W}_3$ module~\eqref{w3module}, and $G_{MN}^{-1}$ is the inverse of the Shapavalov matrix 
\be
G_{MN} = \braket{h_0, q_0, M | h_0,q_0, N}  \; . 
\ee
For general $\Phi_{\alpha_1}$, the matrix elements
\be
\bra{h_0, q_0, M} q^{L_0} \Phi_{\alpha_1}(z) \ket{h_0, q_0, N} \;  
\ee
are not defined uniquely. 
For this reason, we focus on the case where the external field $\Phi_{\alpha_1}(z)$ satisfies~\eqref{alpha4}, from which it follows that  
\begin{equation}
-\frac{h_1^2}{5} = \frac{9 q_1^2}{2} \; .
\end{equation}
This constraint is satisfied when the vector $j_1$ corresponding to the field $\Phi_{\alpha_1}$ is given by $j_1= a w_1$ or $j_1= a w_2$. Without loss of generality, we choose $j_1=aw_1$, which implies 
\begin{equation}
h_1= -a\; , \qquad q_1 = -\frac{i}{3}\sqrt{\frac{2}{5}}a \; .
\label{dimd1} 
\end{equation}

Similarly, the holomorphic $\mathfrak{sl}_3$ one-point conformal block $\mathcal{F}(\alpha, \alpha_1, q)_{\mathfrak{sl}_3}$ is defined by~\eqref{1pt}, but now the states $ \ket{h_0,q_0, N}$ belong to the $\mathfrak{sl}_3$ module, obtained by restricting the set $I$ (\ref{w3module}) to the conditions (\ref{sl3module}). For the fields $\Phi_{\alpha}, \Phi_{\alpha_1}$ with $h$'s and $q$'s given by~\eqref{dimd} and~\eqref{dimd1} respectively, we compute the contribution to the $\mathfrak{sl}_3$ one-point conformal block up to the second level. The matrix elements involved in this computation are listed in appendix~\ref{apA}. Here we present the result
\begin{equation} 
\begin{split}
\mathcal{F}(\alpha, \alpha_1,q)_{\mathfrak{sl}_3} 	& = 1 + \bigg( \frac{18m_1m_2 - a^2(m_1+m_2) - 3a(m_1 + m_2)}{9m_1m_2}\bigg)q \, + \\
            & \quad + \frac{1}{162}\bigg( \frac{(a+6)(a+3)a(a-3)}{m_2 - 1} \, + \\
            & \quad + \frac{\big((a+3)a - 18m_1\big)\big((a+3)a - 36(m_1 - 1)\big)}{m_1(m_1 - 1)} \, + \\
            & \quad - \frac{(a+3)a\big((a+3)a(m_1 - 1) + 36(m_1 + 1)\big)}{(m_1 + 1)m_2} \, + \\
            & \quad + \frac{2(a+3)a(a-3m_1)(a+3m_1 + 3)}{(m_1 + 1)m_1(m_1 + m_2 + 1)}\bigg) q^2 + \dots \; .
\end{split}   
\label{1pt2} 
\end{equation}
As we see from the matrix elements in appendix \ref{apA}, the computation becomes quite tedious, even at the second level. For this reason, it is clear that another method (e.g., the Casimir approach described in~\cite{Kraus:2017ezw,Alkalaev:2022kal}) is needed to compute a general expression of the one-point conformal block. Nevertheless, these first terms of (\ref{1pt2}) will help us to compare with the computation from the Wilson lines approach.

\textbf{Zero-point Conformal Block.} 
 In the case when  $\Phi_{\alpha_1}$ is given by the identity operator and $\Phi_{\alpha}$ is  fully degenerate, the $\mathfrak{sl}_3$ one-point conformal block $\mathcal{F}(\alpha, \alpha_1, q)_{\mathfrak{sl}_3}$ reduces to the $\mathfrak{sl}_3$ character $\chi^{\mathfrak{sl}_3}_j$ of the finite-dimensional representation with highest-weight vector $j= m_1 w_1 + m_2 w_2$. This character can be computed using the Weyl character formula (the details of this computation are given in appendix~\ref{apweylchafor}), 
\begin{equation}
\chi^{\mathfrak{sl}_3}_j=  \frac{q^{-m_1-m_2}(-1+q^{1+m_1})(-1+q^{1+m_2})(-1+q^{2+m_1+m_2})}{(-1+q)^3(1+q)} \; . 
\label{chasl3} 
\end{equation}
\section{One-point conformal blocks from AGT} 
\label{agt1} 
In this section, we move to compute the $\mathcal{W}_3$ one-point conformal block~\eqref{1pt} in the limit $c \rightarrow \infty$ up to the second level of the module~\eqref{w3module}. To this end we use the algorithm based on the AGT correspondence (notations and some details of the algorithm are given in appendix~\ref{apagtrelation}), which leads to the following expression 
\begin{align}
\mathcal{F}(\alpha, \alpha_1,q) 		& = 1 + \bigg(\frac{18m_1 m_2 - a^2(m_1 + m_2) -3a(m_1 + m_2)}{9m_1m_2}\bigg) q \, + \displaybreak[0] \notag \\
            & + \bigg( \frac{(a+3)^2(a-3m_1 -3)(a-3m_2)}{81(m_1 + 1)m_2} \, + \displaybreak[0] \notag \\
            & + \frac{(a+1) \big(a^2(m_1 + m_2) + 3a(m_1 + m_2) - 9(a+3)m_1m_2\big)}{9m_1m_2} \, + \displaybreak[0] \notag \\
            & - \frac{(a+3)^2(a-3m_1)}{27m_1} - \frac{(a+3)^2(a - 3m_2 - 3)}{27(m_2 + 1)} + \frac{1}{2}(a^2 - a - 2) \, + \displaybreak[0] \notag \\
            & +\frac{(a+6)(a+3)(a-3m_1 + 3)(a - 3m_1)}{162m_1(m_1-1)} - \frac{(a+3)(a+3m_2+3)(a-3m_2)}{27m_2(m_2+1)} \, + \displaybreak[0] \notag \\
            & + \frac{(a+6)(a+3)(a-3m_2 + 3)(a-3m_2)}{162 m_2(m_2 - 1)} \, + \displaybreak[0] \notag \\
            & + \frac{(a+3)(a-3m_1)(a + 3m_1 + 3)(a - 3m_1 - 3m_2 - 3)}{81m_1(m_1 + 1)(m_1 + m_2 + 1)} \, + \displaybreak[0] \notag \\
            & \frac{1}{18} (a+6)(a+3) + \frac{a}{3} + 1\bigg) q^2 + \dots \; .
\label{w3cb} 
\end{align}
Now, we propose the following formula that relates the $\mathcal{W}_3$ block to the $\mathfrak{sl}_3$ one in the following way
\begin{equation} 
\frac{\mathcal{F}(\alpha, \alpha_1,q)}{\mathcal{F}(\alpha, \alpha_1,q)_{\mathfrak{sl}_3}} = \frac{\chi^{\mathcal{W}_3}}{\chi^{\mathfrak{sl}_3}} \; ,
\label{relation} 
\end{equation}
where $\chi^{\mathcal{W}_3}$ and $\chi^{\mathfrak{sl}_3}$ are $\mathcal{W}_3$ and $\mathfrak{sl}_3$ (for infinite-dimensional representations) characters respectively, given by

\begin{equation}
\begin{split}
\chi^{\mathcal{W}_3} 	& = \frac{1}{\prod_{i=1}^{\infty} (1-q^i)^2}\; , \\ 
\chi^{\mathfrak{sl}_3}	& = \frac{1}{(1-q)^3 (1+q)}\; ,
\end{split} 
\end{equation}
where the second line is obtained by fixing the proper normalization of~\eqref{chasl3} consistent with the standard normalization of CB's and then taking the limit ($m_1, m_2 \rightarrow \infty$). By substituting~\eqref{w3cb} in~\eqref{relation} and solving for $\mathcal{F}(\alpha, \alpha_1,q)_{\mathfrak{sl}_3}$, we recover the expression~\eqref{1pt2}. This confirms, up to the second level, our proposal. The motivation of this conjectured formula comes from the fact that a similar relation~\cite{Alkalaev:2016fok} holds for the Virasoro block $\mathcal{F}_{\text{Vir}}$ in the large central charge limit and $\mathfrak{sl}_2$ block, namely 
\be
\frac{\mathcal{F}_{\text{Vir}}}{\mathcal{F}_{\mathfrak{sl}_2}} = \frac{\chi^{\text{Vir}}}{\chi^{\mathfrak{sl}_2}}= \frac{1-q}{\prod_{i=1}^{\infty} (1-q^i)}\;. 
\ee
 The relation~\eqref{relation} allows effectively to compute higher-level contributions to the $\mathfrak{sl}_3$ block. Note that the characters and blocks involved in~\eqref{relation} refer to infinite-dimensional representations. Nevertheless, we can still use this formula 
 for finite-dimensional representations, truncating the series up to an appropriate level.

\section{Conformal Blocks through Wilson line operators} 
\label{wilsonline} 
In \cite{Alkalaev:2020yvq}, it was proved that the expectation value of the Wilson network operator computes the $\mathfrak{sl}_2$ toroidal one-point conformal block. This section aims to show that a similar result is obtained in the case of $\mathfrak{sl}_3$ algebra. 

In the case of the spherical topology and the $\mathfrak{sl}_3$ algebra, in  \cite{Besken:2016ooo}, it was given the Wilson lines description of the $\mathfrak{sl}_3$ four-point block. Here we will combine the ideas of \cite{Alkalaev:2020yvq} and \cite{Besken:2016ooo} to proceed with our task. Before introducing the object we start with some definitions and notations. In the subsequent, we will work with finite-dimensional representations $\mathcal{R}$ of $\mathfrak{sl}_3$ that, in general, are characterized by a highest-weight vector $j$ given by \eqref{alpha2}. 
We will label the finite-dimensional representations $\mathcal{R}$ of $\mathfrak{sl}_3$ by the vectors $\alpha, \alpha_1$, having in mind that they correspond to highest-weight vectors $j, j_1$ according to~\eqref{alpha2} and~\eqref{alpha3}. The Wilson line operators in thermal $\AdS_3$ are given by 
\begin{equation} 
W_{\alpha}[x,y]=\mathcal{P}\exp \bigg(-\int_{x}^{y} \a \bigg) = \exp \left( (x-y) (   L_1+\frac{1}{4} L_{-1}  ) \right) \; ,
\label{wlosl3} 
\end{equation}
where $L_1, L_{-1}$ are defined in the representation $\mathcal{R}_{\alpha}$ of the $\mathfrak{sl}_3$ algebra.

The \emph{Wilson network operator} allows to compute the conformal block and contains a number of elements   (see, e.g.,~\cite{Alkalaev:2020yvq}) that are listed below for completeness. 
\begin{enumerate}
\item  $\Phi_{(h_{\alpha}, q_{\alpha})} (z_{\alpha}, \bar{z}_{\alpha})$ on the boundary is attached to a bulk-to-boundary Wilson line operator $W_{\alpha}[z_b, z_{\alpha}]$ acting in the representation $\varR_{\alpha}$, which connects the boundary point $z_{\alpha}$ to the bulk point $z_b$. 
\item Bulk-to-bulk Wilson line operator $W_{\beta}[z_1, z_2]$ in the representation $\varR_{\beta}$, which connects two bulk points and acts trivially for a contractible loop. In the toroidal topology, there exists a non-trivial bulk-to-bulk Wilson loop operator $W_{\beta}[z_b, z_b + 2\pi \tau]$, associated with a non-contractible cycle.
\item A vertex in the bulk, obtained when three Wilson line operators, associated with the representations $\varR_{\alpha}$, $\varR_{\beta}$ and $\varR_{\gamma}$, meet each other in the same point. This vertex is described by the trivalent intertwining operator 
\be
I_{\alpha; \beta, \gamma} \, : \quad \varR_{\beta} \otimes \varR_{\gamma} \; \longrightarrow \varR_{\alpha} \; , 
\label{eqn:WL3} 
\ee
defined by the invariance property 
\be
I_{\alpha; \beta, \gamma} U_{\beta} U_{\gamma} = U_{\alpha} I_{\alpha; \beta, \gamma} \; , 
\ee
where $U_a$, ($a=\alpha, \beta, \gamma$), are  elements of the gauge group. 
\end{enumerate}

With a proper choice of the gauge, we can simplify the expression of the Wilson loop operator~\cite{Alkalaev:2020yvq}. Due to the gauge covariance of the Wilson line operator, we can perform a gauge transformation of the connection
\be
\Omega = U \tilde{\Omega} U^{-1} \; , 
\ee
where, if we choose as gauge element
\be
U =  e^{-\frac{i}{2}L_{-1}}e^{i L_{1}} e^{-i\frac{\pi}{2}L_0} \; , 
\label{eqn:WL2} 
\ee
we obtain 
\be
\tilde{\Omega} = -i L_0 \, d z \; .
\ee
With this particular choice, known as \emph{diagonal gauge}, we have that the Wilson loop reads
\be
W_i [0, 2\pi \tau] = e^{2\pi i \tau L_0} = q^{L_0} \qquad \text{ with } \, q = e^{2\pi i \tau} \; . 
\label{eqn:WL4} 
\ee
The only element that is affected by this choice of the gauge is the external field, which has to transform accordingly. We consider as the boundary state the lowest-weight state of the representation that has to be transformed as 
\be
\ket{lw}_{\alpha} \; \longrightarrow \; \ket{\tilde{lw}}_{\alpha} \equiv U_{\alpha}^{-1} \ket{lw}_{\alpha} \; , 
\ee
where $U_{\alpha}$ is the gauge group element~\eqref{eqn:WL2} in the representation $\varR_{\alpha}$.

\subsection{Computation of the Wilson line operators} \label{cwlo} 
We are interested in performing the computation of the Wilson network operator exploiting the tensor products of representations of the $\sl_3$ algebra. This is based on the fact that every state in a given representation of the algebra can be represented as a symmetric traceless tensor. Let us show how this procedure works.

The fundamental representation of $\mathfrak{sl}_3$ is characterized by the highest-weight vector $j= w_1= (1,0)$. This representation contains three states given by~\eqref{voffundamentalr}, which for convenience, we denote as
\begin{equation} 
\ket{e_1} = w_1, \quad \ket{e_2} = w_2-w_1, \quad \ket{e_3} = -w_2 \; .
\label{fundstates} 
\end{equation}
Similarly, we denote the states of the anti-fundamental representation $(0,1)$ (the conjugated of the fundamental) 
\begin{equation} 
\ket{\bar{e}^1} = -w_1, \quad \ket{\bar{e}^2} = w_1-w_2, \quad \ket{\bar{e}^3} = w_2 \; .
\label{conjustates}
\end{equation}
For a generic $\mathfrak{sl}_3$ representation with a highest-weight vector 
\be
j = m_1 w_1+m_2 w_2 \equiv (m_1, m_2) \; ,
\ee 
the whole vector space of the representation is obtained by applying the lowering generators $L_1$, $W_1$, $W_2$ to the highest-weight vector. Each of these states can be written as the tensor product of the states of the fundamental representation~\eqref{fundstates} and its conjugate~\eqref{conjustates} in the following way
\begin{equation}
\begin{split}
T^{k_1....k_{m_2}}_{i_1...i_{m_1}} 	& = \frac{1}{m_1! m_2!} \ket{e_{(i_1}} \otimes \ket{e_{i_2}} \otimes \cdots \otimes \ket{e_{i_{m_1})}}  \, \ket{\bar{e}^{(k_1}} \otimes \ket{\bar{e}^{k_2}} \otimes \cdots \otimes \ket{\bar{e}^{k_{m_2})}} = \\
							& = \frac{1}{m_1! m_2!} \ket{e_{(i_1}...e_{i_{m_1})}  \bar{e}^{(k_1}...\bar{e}^{k_{m_2})}} \; ,  
\end{split}
\label{sytensor} 
\end{equation}
where the parentheses of lower and upper indices denote the symmetrization, and the tensor $T^{k_1....k_{m_2}}_{i_1...i_{m_1}} $ has to be traceless. The highest-weight vector corresponds to the case when all indices  $i$ are equal to 1 for the fundamental representation, while for the anti-fundamental one, we require all the indices $k$ to be equal to $3$. Below, we will need to compute the tensor product of the kind $\mathcal{R}_{\alpha} \otimes  \mathcal{R}_{\alpha_1}$. In particular, here we need to extract the representation $\mathcal{R}_{\alpha}$, i.e., we need the projection 
\begin{equation}  
P: \mathcal{R}_{\alpha} \otimes  \mathcal{R}_{\alpha_1} \rightarrow  \mathcal{R}_{\alpha} \; .
\label{defofprojector} 
\end{equation}
The role of this projector is played by the intertwining operator $I_{\alpha; \alpha, \alpha_1}$~\eqref{eqn:WL3}. In the way this intertwining operator is defined, its matrix elements are given by the Clebsch-Gordan  coefficients, i.e. 
\be 
\bra{a} I_{\alpha; \alpha, \alpha_1} \ket{b} \otimes \ket{c} = C_{abc} \; ,
\ee
where $a, b \in \mathcal{R}_{\alpha}$, $c \in  \mathcal{R}_{\alpha_1}$, and $C_{abc}$ is the Clebsch-Gordan coefficient. However, in what follows, we are going to define this projector explicitly, since for $\mathfrak{sl}_3$, the Clebsch-Gordan coefficients are not known for generic representations. 

We want to show that the following expression holds, where $\mathcal{F}(\alpha, \alpha_1,q)_{\mathfrak{sl}_3}$ is the $\mathfrak{sl}_3$ one-point conformal block introduced in~\eqref{1pt2}
\begin{equation}
\begin{split}
V_{\alpha | \alpha_1}(\tau) \deffl \sum_{\ket{\alpha} \in \mathcal{R}_{\alpha}} \bra{\alpha} W_{\alpha}[z_b, z_b+2\pi \tau] & I_{\alpha;\alpha, \alpha_1} \ket{\alpha} \otimes W_{{\alpha}_1}[z_b, z_1] \ket{lw}_{\alpha_1} =\\&=
\tilde{C}(\alpha,\alpha_1)  \mathcal{F}(\alpha, \alpha_1,q) _{\mathfrak{sl}_3} \; ,
\label{w1pt} 
\end{split}
\end{equation}
where $z_1$ is the point on the boundary of the solid torus that is a geometric representation of the thermal $\AdS_3$, $z_b$ is an arbitrary point in the bulk of $\AdS_{3}$, $\ket{lw}_{\alpha_1}$ is the lowest-weight state of the representation $\mathcal{R}_{\alpha_1}$ with $j_1 = (a,0)$ and $\tilde{C}(\alpha,\alpha_1)$ is an overall constant irrelevant to our discussion. By using the diagonal gauge~\eqref{eqn:WL4} it can be shown that $V_{\alpha|\alpha_1}(\tau)$ does not depend on $z_b$ and $z_1$, and it can be expressed as
\begin{equation} 
V_{\alpha | \alpha_1}(\tau) = Tr_{\alpha}\, \bra{\alpha} q^{L_0} I_{\alpha; \alpha, \alpha_1} \ket{\alpha} \otimes U_{\alpha_1}^{-1} \ket{lw}_{\alpha_1} \; , 
\label{w1pt2} 
\end{equation}
where $Tr_{\alpha}= \sum_{\ket{\alpha} \in \mathcal{R}_{\alpha}}$ and $U_{\alpha_1}^{-1} $ is given by~\eqref{eqn:WL2} in the representation $\mathcal{R}_{\alpha_1}$. Now, properly inserting resolutions of identities
\be
\Id = \sum_{\varR_\beta} \ket{\beta}\bra{\beta} \; , 
\ee
where $\sum_{\varR_\beta}$ stands for the sum over all the states in the representation $\varR_{\beta}$, we can write, in a symbolic way,
\be
V_{\alpha|\alpha_1}(\tau) = \sum_{\varR_\alpha}\sum_{\varR_\gamma} \bra{\alpha} q^{L_0}\ket{\alpha}\Big( \bra{\alpha}I_{\alpha;\alpha, \gamma} \ket{\alpha}\otimes \ket{\gamma}\Big) \bra{\gamma}U^{-1}_{\alpha_1}\ket{lw}_{\alpha_1} \; . 
\label{eqn:WLC1}
\ee
Let us see how to represent each of the factors in~\eqref{eqn:WLC1}.
\begin{enumerate}
\item The first term can be represented in terms of the so-called \textit{Wigner D-matrix}. We will be interested in particular in the fundamental and anti-fundamental representations, where the $D$-matrices read respectively  
\be
D_{ij} \deffl \bra{e_i} q^{L_0}\ket{e_j} = \begin{pmatrix} q & 0 & 0 \\ 0 & 1 & 0 \\ 0 & 0 & q^{-1} \end{pmatrix}  \; , \qquad %
\bar{D}^{ij} \deffl \bra{\bar{e}^i} q^{L_0}\ket{\bar{e}^j} = \begin{pmatrix} q^{-1} & 0 & 0 \\ 0 & 1 & 0 \\ 0 & 0 & q \end{pmatrix}  \; .
\label{eqn:WLC2}
\ee
\item The second factor in~\eqref{eqn:WLC1}, the one involving the intertwining operator, acts as a projector. Its explicit action will be explained later.
\item The third term consists in the coordinates of the transformed lowest-weight state of the external representation. Again, focusing on the fundamental representation, we have that the transformed lowest-weight state is\footnote{In the fundamental representation, we use the following forms of the generators of $\mathfrak{sl}_2$ subalgebra of $\mathfrak{sl}_3$\be
L_{1} = \sqrt{2}\begin{pmatrix} 					0 	& 0 	& 0 \\
										1	& 0	& 0 \\
										0	& 1	& 0
			\end{pmatrix} \; , \quad 
 L_0 = \begin{pmatrix}	 					1	& 0	& 0 \\
										0	& 0 	& 0 \\
										0	& 0	& -1 
			\end{pmatrix} \; , \quad 
L_{-1} = \sqrt{2} \begin{pmatrix}					0	& -1	& 0 \\
										0	& 0	& -1 \\
										0	& 0	& 0
			\end{pmatrix} \; .
\ee}
\be
U^{-1}_{\footnotesize \text{fund}} \ket{e_3} \propto \ket{e_1} + \sqrt{2}\ket{e_2} + \ket{e_3} \; .
\ee
This means that, in general, defining 
\be
p_k \deffl \bra{e_k}U^{-1}_{\footnotesize \text{fund}} \ket{e_3} = \delta_{k,1} + \sqrt{2}\delta_{k,2} + \delta_{k,3} \qquad k = 1,2,3 \; , 
\label{eqn:WLC3}
\ee
we have that 
\be \label{fundamatrix2}
\bra{ e_{k_1} e_{k_2} \cdots e_{k_a}}U^{-1}_{\alpha_1} \ket{lw}_{\alpha_1} \sim p_{k_1} p_{k_2} \cdots p_{k_a} \; . 
\ee
\end{enumerate}
At this stage, we state the main result of this section. Even though we do not have, as of this writing, a general proof of~\eqref{w1pt} for generic representations $\mathcal{R}_{\alpha}, \mathcal{R}_{\alpha_1}$, we have tested it for some non-trivial representations given by the following values 
\begin{equation} 
(m_1, m_2, a)= \{  (2,2,3), (2,2,6), (2,3,3), (2,3,6), (3,3,3), (3,3,6), (4,4,3), (4,4,6)\} \; ,
\label{ourexamples} 
\end{equation}
obtaining that in all these cases,~\eqref{w1pt} holds precisely.

Let us anticipate here that, to obtain a non-vanishing one-point toroidal conformal block, we have to require $a \equiv 0 \mod 3$. The details of the case ($2,2,3$) are given in appendix~\ref{wilexamples} as an example. Furthermore, as a complementary motivation of~\eqref{w1pt} for the case when $a=0$ (i.e., when $\mathcal{R}_{\alpha_1}$ is the trivial representation), we show explicitly that for representations $\mathcal{R}_{\alpha}$ with highest-weight vector $j=(m_1, 0)$ or $j= (0,m_2)$, $V_{\alpha |\alpha_1}$ in~\eqref{w1pt} is equal to the character~\eqref{chasl3}. 

\subsubsection{Zero-point conformal block} \label{zpcb} 
The simplest case of~\eqref{w1pt}  is when $\mathcal{R}_{\alpha_1}$ is the trivial representation (one-dimensional), in which case we expect to find the character~\eqref{chasl3} (or \textit{zero-point conformal block}) of the $\mathfrak{sl}_3$ representation. Let us check this statement. 

We show this by choosing as highest-weight vector of $\mathcal{R}_{\alpha}$ the vector $j = (m_1,0)$\footnote{The same procedure holds in the case $j = (0,m_2).$}. Since $\mathcal{R}_{\alpha_1}$ is the trivial representation, the intertwining operator $I_{\alpha; \alpha, 0}$ (or the projector~\eqref{defofprojector}) reduces to the identity operator, thus 
\begin{equation}
\begin{split}
V_{\alpha|0}(\tau) \equiv V_{\alpha}  	& =  \sum _{i_1, \dots, i_{m_1}= 1,2,3}   \bra{e_{i_1} \dots e_{i_{m_1}}} q^{L_0} \ket{e_{(i_1} \dots e_{i_{m_1)}}} =  \\ 
							& = \sum_{i_1, \dots, i_{m_1} = 1,2,3}  D_{i_1( i_1} \dots D_{i_{m_1} i_{m_1})} \; .
\label{wch1} 
\end{split}
\end{equation}
Due to the symmetrization of the indices, the above expression can be written as 
\begin{equation}
V_{\alpha} = \sum_{\substack{m,n,k=0 \\ m+n+k=m_1}}^{m_1}q^{m}q^{-n}(1)^k \; ,
\label{wch2} 
\end{equation} 
and after some algebra, we can show that~\eqref{wch2} gives 
\begin{equation} 
V_{\alpha} =  \sum _{r=1}^{m_1} \left(q^{-r}+q^r\right) \left(-M(m_1+r)+\frac{m_1+r}{2}-r+1\right)-M(m_1)+\frac{m_1}{2}+1 \; ,
\label{wch3}
\end{equation}
where $M(r)= \frac{1-(-1)^r}{4}$. By expanding $\chi^{\mathfrak{sl}_3}_{j=(m_1,0)}$ of~\eqref{chasl3} in $q$, we can see that this character is equal to~\eqref{wch3}. 

We would like to obtain the same result in the general case when $m_2 \neq 0$, i.e., we want to show the following equation  
\begin{equation}
\sum_{\substack{i_1, \dots , i_{m_1}= 1,2,3 \\ k_1, \dots , k_{m_2}=1,2,3}}  \bra{\bar{e}^{k_{1}} \dots \bar{e}^{k_{m_2}} \, e_{i_{1}} \dots e_{i_{m_1}} } q^{L_0} T^{k_1\dots k_{m_2}}_{i_1...i_{m_1}} =  \chi^{\mathfrak{sl}_3}_j\; , 
\label{chfromtensor} 
\end{equation}
where $\chi^{\mathfrak{sl}_3}_j$  is the character~\eqref{chasl3}, and the tensor $T^{k_1....k_{m_2}}_{i_1...i_{m_1}}$ is the symmetric (in lower and upper indices)  and traceless tensor from~\eqref{sytensor}. We do not have a general proof of~\eqref{chfromtensor}. However, the numerical results we have obtained for many values of $j$ confirm this equation.

\subsubsection{One-point conformal block}
In this section, we elaborate on the expression~\eqref{w1pt2} (which we claimed that reproduces the $\mathfrak{sl}_3$ one-point conformal block $ \mathcal{F}(\alpha, \alpha_1,q)_{\mathfrak{sl}_3})$ for general representations $\varR_{\alpha}$ and $\varR_{\alpha_1}$ labelled by the highest-weight vectors $j = (m_1, m_2)$ and $j_1 = (a, 0)$, respectively. The purpose is to simplify this expression in order to compare it with $ \mathcal{F}(\alpha, \alpha_1,q)_{\mathfrak{sl}_3}$.
In what follows, we will not give too much attention to overall constants since they are generally ineffective. For this reason, some of the following equalities have to be considered to hold up to a constant.

The idea is to represent~\eqref{w1pt2} as a symmetric traceless tensor. The first step will be constructing a tensor with the right structure to represent the one-point block but without the requirements of symmetry and tracelessness, which will be imposed at the end. So, let us start defining the following tensor 
\begin{multline}
^{b_1 \cdots b_{m_2}}_{a_1 \cdots a_{m_1}}T^{r_1 \cdots r_{m_2}}_{s_1 \cdots s_{m_1}; k_1 \cdots k_a} \deffl \\
= \bra{\bar{e}^{r_1} \cdots \bar{e}^{r_{m_2}} \, e_{s_1} \cdots e_{s_{m_1}}} q^{L_0} \ket{e_{a_1} \cdots e_{a_{m_1}} \bar{e}^{b_1} \cdots \bar{e}^{b_{m_2}}} \, \bra{e_{k_1} \cdots e_{k_a}}U^{-1}_{\alpha_1}\ket{lw}_{\alpha_1} \, , 
\label{eqn:OCB1} 
\end{multline}
which, thanks to the definitions~\eqref{eqn:WLC2},~\eqref{eqn:WLC3} and~\eqref{fundamatrix2}, can be rewritten as 
\be
^{b_1 \cdots b_{m_2}}_{a_1 \cdots a_{m_1}}T^{r_1 \cdots r_{m_2}}_{s_1 \cdots s_{m_1}; k_1 \cdots k_a} = D_{a_1 s_1} \cdots D_{a_{m_1}s_{m_1}} \bar{D}^{b_1 r_1} \cdots \bar{D}^{b_{m_2} r_{m_2}} p_{k_1} \cdots p_{k_a} \; . 
\label{eqn:OCB2} 
\ee
The tensor defined in~\eqref{eqn:OCB1} resembles the quantity~\eqref{w1pt2} we are interested in. However, we need to insert the intertwining operator. In order to do that, let us recall that it acts as a projector~\eqref{defofprojector}, i.e., it translates in the insertion of a tensor $P$, which allows us to define  
\be
^{b_1 \cdots b_{m_2}}_{a_1 \cdots a_{m_1}}M^{c_1 \cdots c_{m_2}}_{d_1 \cdots d_{m_1}} = \big( P^{c_1 \cdots c_{m_2}}_{d_1 \cdots d_{m_1}}\big)^{s_1 \cdots s_{m_1} ; k_1 \cdots k_a}_{r_1 \cdots r_{m_2}} \; ^{b_1 \cdots b_{m_2}}_{a_1 \cdots a_{m_1}}T^{r_1 \cdots r_{m_2}}_{s_1 \cdots s_{m_1}; k_1 \cdots k_a} \; . 
\label{eqn:OCB3} 
\ee
In order to construct the tensor $P$, we follow the method described in~\cite{Besken:2016ooo}. It has to be a product of $\sl_3$ invariant tensors, which are
\be
\delta^{i}_j \; , \quad \epsilon^{ijk} \; , \quad \epsilon_{ijk} \; .
\ee
However, since the final result has to be traceless and symmetric, the only possible contractions we are interested in are with
\be
\delta^k_r \; , \qquad \epsilon^{csk} \; , 
\ee
where we have labelled the indices coherently to~\eqref{eqn:OCB3}. If we suppose that $P$ is made out of $x$ contractions with $\delta^k_r$ and $y$ contractions with $\epsilon^{csk}$, since the remaining indices have to be the same in number as the expected ones, i.e., $m_1$ lower and $m_2$ upper indices, a simple counting shows us that $a$ has to satisfy the condition
\be
a = 3l \; , \qquad l \in \ZZ_{\ge 0} \; .
\ee
In this way, we obtain that the tensor that represents the intertwining operator is 
\begin{multline}
\big( P^{c_1 \cdots c_{m_2}}_{d_1 \cdots d_{m_1}}\big)^{s_1 \cdots s_{m_1} ; k_1 \cdots k_a}_{r_1 \cdots r_{m_2}} = \delta^{s_1}_{d_1} \cdots \delta^{s_{m_1 - l}}_{d_{m_1 - l}} \delta^{k_1}_{d_{m_1 - l + 1}} \cdots \delta^{k_l}_{d_{m_1}} \delta^{c_1}_{r_1} \cdots \delta^{c_{m_2 - l}}_{r_{m_2 - l}} \times \\
\times \delta^{k_{l+1}}_{r_{m_2 - l +1}} \cdots \delta^{k_{2l}}_{r_{m_2}} \epsilon^{c_{m_2 -l + 1} s_{m_1 - l + 1}k_{2l + 1}} \cdots \epsilon^{c_{m_2} s_{m_1} k_{3l}} \; .  \label{seconddefofP}
\end{multline}
Inserting this projector into~\eqref{eqn:OCB3}, we end up with 
\begin{multline}
^{b_1 \cdots b_{m_2}}_{a_1 \cdots a_{m_1}}M^{c_1 \cdots c_{m_2}}_{d_1 \cdots d_{m_2}} = D_{a_1d_1} \cdots D_{a_{m_1-l}d_{m_1-l}} \, \bar{D}^{b_1 c_1}\cdots \bar{D}^{b_{m_2 - l} c_{m_2 - l}} p_{d_{m_1 - l + 1}} \cdots p_{d_{m_1}} \times \\
\times \bar{p}^{b_{m_2 - l + 1}} \cdots \bar{p}^{b_{m_2}} \, A^{c_{m_2 - l + 1}}_{a_{m_1 - l + 1}} \cdots A^{c_{m_2}}_{a_{m_1}} ,
\label{eqn:OCB4} 
\end{multline}
where we have introduced the new quantities 
\be
\bar{p}^{b} \deffl \bar{D}^{br}p_r \; , \qquad A^{c}_{a} \deffl \epsilon^{csk} D_{as} p_k \; .
\ee
We want to make this tensor symmetric, and we do this procedure just formally, remembering in what follows that we have to consider all the possible cyclic permutations of the indices. Hence we write 
\be \label{thesymmetrictensorM}
^{b_1 \cdots b_{m_2}}_{a_1 \cdots a_{m_1}}M^{c_1 \cdots c_{m_2}}_{d_1 \cdots d_{m_1}} \equiv \; ^{b_1 \cdots b_{m_2}}_{a_1 \cdots a_{m_1}}M^{(c_1 \cdots c_{m_2})}_{(d_1 \cdots d_{m_1})} \; . 
\ee
The last step consists in making this tensor traceless. We do this by exploiting the procedure described in~\cite{Besken:2016ooo}, where it is shown that a symmetric tensor can be made traceless as
\be
^{b_1 \cdots b_{m_2}}_{a_1 \cdots a_{m_1}}\tilde{M}^{c_1 \cdots c_{m_2}}_{d_1 \cdots d_{m_1}} = \sum_{n=0}^{\min} \varC_n \delta^{(c_1}_{(d_1} \cdots \delta^{c_n}_{d_n} \;^{b_1 \cdots b_{m_2}}_{a_1 \cdots a_{m_1}}M^{c_{n+1} \cdots c_{m_2})f_1 \cdots f_n}_{d_{n+1} \cdots d_{m_1})f_1 \cdots f_n} \; , 
\label{eqn:OCB5} 
\ee
where
\be
\varC_n = (-1)^n \frac{(m_1 + m_2 - n + 1)!}{(m_1+m_2+1)!(m_1-n)!(m_2-n)! n!} \; , \qquad \min = \min(m_1, m_2) \; , 
\ee
and in each summand, we take the sums over repeated indices $f_i$. 

Finally,~\eqref{w1pt} can be written as
\begin{equation}
 {}_{a_1\cdots a_{m_1}}^{b_1 \cdots b_{m_2}}\tilde{M}_{(d_1 \cdots d_{m_1})}^{(c_1 \cdots c_{m_2})}\delta_{b_1 c_1} \cdots \delta_{b_{m_2} c_{m_2}}\delta^{a_1 d_1} \cdots \delta^{a_{m_1} d_{m_1}} = \tilde{C}(\alpha,\alpha_1) \mathcal{F}(\alpha, \alpha_1,q)_{\mathfrak{sl}_3}\; . 
\label{mainequation} 
\end{equation}
As of the date of this work, we have not found a general proof of~\eqref{mainequation}, but many numerical computations (which are partially described in the appendix~\ref{wilexamples}) confirm this equality. 

%
\section{Discussion} 
\label{conclusion} 
In this work we studied the semiclassical limit of $\varW_3$ CFT on the torus, with a focus on the 
$\mathfrak{sl}_3$ one-point conformal blocks of fully degenerate primary fields for an external field satisfying~\eqref{alpha4}. We compute the $\varW_3$ block in the light  $c\rightarrow \infty$ limit using AGT correspondence. In order to obtain the $\sl_3$ block we propose the relation~\eqref{relation}. The utility of this relation is twofold since it serves as a check of the dual representation for $\mathfrak{sl}_3$ conformal block and provides a computational method to calculate its higher-level contributions.  

The dual description of $\mathfrak{sl}_3$ one-point toroidal conformal block is given in terms of the Wilson lines in $3d$ Chern-Simons gravity. The proposal summarizes in~\eqref{mainequation}. The dual approach seems to give more manageable computations compared to those based on the original definition, at least in its construction. The simplest test of this proposal is the case when the one-point block  reduces to the zero-point block, which we confirmed analytically and numerically in section~\ref{zpcb}. Further, we have performed more general tests for specific cases of representations~\eqref{ourexamples}, obtaining that all these examples confirm~\eqref{w1pt}. The absence of a general expression of $\mathfrak{sl}_3$ Clebsch-Gordan coefficients forces us to use the tensor technique in section~\ref{cwlo}, the general proof of~\eqref{mainequation} remains an open problem. 

From the Wilson line construction a natural generalization follows, that is the study of the dual description in the cases of higher-point conformal blocks, which show additional features, like the presence of different intermediate channels (usually called $s$ and $t$), as shown for the $\sl_2$ case in~\cite{Alkalaev:2020yvq}. This generalization is straightforward in the context of the Wilson lines formulation of the conformal blocks, according to the construction of section~\ref{wilsonline}. Also, it would be worth to extend this approach to the whole $\varW_3$ CFT, which would require to focus also on quantum corrections (for related consideration see~\cite{Hikida:2018dxe}).   

Another interesting  question is how the Wilson line approach works if multiplicities  in the representations under consideration are present. From the dual construction it follows that in such cases we can obtain multiple nonequivalent expressions for the projector~\eqref{defofprojector}. For instance, if the external field is in the 
adjoint representation, there is a multiplicity two for the conformal blocks in the $\mathcal{W}_3$ CFT~\cite{Belavin:2016qaa}, which corresponds to the fact that Clebsch-Gordan coefficients are not uniquely defined~\cite{Alex:2010wi}. In this case, the multiplicity is present also in the definition of the projector. It would be interesting to establish a precise correspondence between uncertainty related to the OPE multiplicities for conformal blocks  
and multiplicities arising in the definition of the projector in the dual construction.

\appendix 
\section{Matrix Elements} 
\label{apA} 
Here, we give the matrix elements that enter in the computation of the $\mathfrak{sl}_3$ conformal block~\eqref{1pt2} at the first and second levels of $\mathfrak{sl}_3$ module. We use the algorithm described in \cite{Kanno:2010kj} to compute these matrix elements. 

\textbf{First level.} From the $\sl_3$ commutation relations, one can find the following products
\begin{equation}
\begin{split}
& \langle  h_0, q_0|\bL_1 \bL_{-1}   |h_0, q_0 \rangle= 2h_0 \; ,\\ 
& \langle  h_0, q_0|\bW_1 \bL_{-1}   |h_0, q_0 \rangle = 3q_0 \; ,   \\
& \langle  h_0, q_0|\bW_1 \bW_{-1}   |h_0, q_0 \rangle = -h_0/5 \; ,
\end{split}
\end{equation}
thus the Shapavalov matrix $G$ at the first level is
\begin{equation}
G      =    
\left(
\begin{array}{cc}
 2 h_0 & 3 q_0 \\
 3 q_0 & -\frac{h_0}{5} \\
\end{array}
\right).
\end{equation}
The matrix elements at the first level are\footnote{All the matrix elements are proportional to $ \langle  h_0, q_0|\Phi_{\alpha_1}|h_0, q_0 \rangle$. But, for the sake of brevity, we omit this factor.}
\begin{equation}
\begin{split}
& \langle  h_0, q_0|\bL_1 \Phi_{\alpha_1} \bL_{-1}   |h_0, q_0 \rangle=  2 h_0+\left(h_1-1\right) h_1,\\ 
& \langle  h_0, q_0|\bW_1 \Phi_{\alpha_1} \bL_{-1}   |h_0, q_0 \rangle =  3 q_0+\frac{1}{2} \left(h_1-1\right) q_1,   \\
&  \langle  h_0, q_0|\bL_1  \Phi_{\alpha_1}\bW_{-1}   |h_0, q_0 \rangle = 3 q_0-\frac{1}{2} \left(h_1-1\right) q_1 ,\\
& \langle  h_0, q_0|\bW_1\Phi_{\alpha_1} \bW_{-1}   |h_0, q_0 \rangle =  \frac{1}{5} \left(h_1-h_0\right)-\frac{\left(h_1-3\right) q_1^2}{4 h_1}.
\end{split}
\end{equation}

\textbf{Second level.} 
At this level, we consider the basis 
\be
\bL_{-1}^2\; , \quad  \bW_{-1}^2\; , \quad \bL_{-1}\bW_{-1}\; , \quad  \bW_{-2} \; .
\ee
We obtain the following products 
\begin{equation}
\begin{split}
& \langle  h_0, q_0|\bL^2_1 \bL^2_{-1}   |h_0, q_0 \rangle= 4 h_0 \left(2 h_0+1\right),\\ 
& \langle  h_0, q_0|\bW^2_1 \bL^2_{-1}   |h_0, q_0 \rangle = 18 q_0^2-\frac{6 h_0}{5} ,   \\
&  \langle  h_0, q_0|\bW_1 \bL_1 \bL^2_{-1}   |h_0, q_0 \rangle = 6 \left(2 h_0+1\right) q_0,\\
& \langle  h_0, q_0|\bW_2 \bL^2_{-1}   |h_0, q_0 \rangle =  12 q_0 ,\\
& \langle  h_0, q_0|\bW^2_1 \bW^2_{-1}   |h_0, q_0 \rangle = \frac{1}{25} h_0 \left(2 h_0+1\right) ,   \\
&  \langle  h_0, q_0|\bW_1 \bL_1 \bW^2_{-1}   |h_0, q_0 \rangle = \frac{1}{5} (-3) \left(2 h_0+3\right) q_0,\\
& \langle  h_0, q_0|\bW_2 \bW^2_{-1}   |h_0, q_0 \rangle =  \frac{6 q_0}{5},  \\
&  \langle  h_0, q_0|\bW_{1}\bL_1\bL_{-1} \bW_{-1}   |h_0, q_0 \rangle = 9 q_0^2-\frac{2}{5} h_0 \left(h_0+1\right) ,\\
& \langle  h_0, q_0|\bW_2 \bL_{-1} \bW_{-1} |h_0, q_0 \rangle =  -\frac{1}{5} \left(4 h_0\right), \\
&  \langle  h_0, q_0|\bW_2  \bW_{-2} |h_0, q_0 \rangle =   \frac{8 h_0}{5}. 
\end{split}
\end{equation}
Hence the Shapavalov matrix $G$ at the second level is
\begin{equation}
G  =\left(
\begin{array}{cccc}
 4 h_0 \left(2 h_0+1\right) & 18 q_0^2-\frac{6 h_0}{5} & 6 \left(2 h_0+1\right) q_0 & 12 q_0 \\
 18 q_0^2-\frac{6 h_0}{5} & \frac{1}{25} h_0 \left(2 h_0+1\right) & \frac{1}{5} (-3) \left(2 h_0+3\right) q_0 & \frac{6 q_0}{5} \\
 6 \left(2 h_0+1\right) q_0 & \frac{1}{5} (-3) \left(2 h_0+3\right) q_0 & 9 q_0^2-\frac{2}{5} h_0 \left(h_0+1\right) & -\frac{1}{5} \left(4 h_0\right) \\
 12 q_0 & \frac{6 q_0}{5} & -\frac{1}{5} \left(4 h_0\right) & \frac{8 h_0}{5} \\
\end{array}
\right).
\end{equation}
The matrix elements at the second level are
\begin{equation}
\begin{split}
& \langle  h_0, q_0|\bL^2_1 \Phi_{\alpha_1} \bL^2_{-1}   |h_0, q_0 \rangle=   8 h_0^2+\left(8 \left(h_1-1\right) h_1+4\right) h_0+\left(h_1-1\right) h_1 \left(\left(h_1-1\right) h_1+2\right),\\ 
&\begin{split} \langle  h_0, q_0|\bL^2_1 \Phi_{\alpha_1} \bW^2_{-1}   |h_0, q_0 \rangle & =    18 q_0^2-6 \left(h_1-1\right) q_1 q_0+\frac{\left(h_1^3-7 h_1+6\right) q_1^2}{4 h_1}   \\& -    \frac{1}{5} \left(h_1-1\right) h_1 \left(h_1+4\right)-\frac{6 h_0}{5}, \end{split}   \\
&  \begin{split} \langle  h_0, q_0|\bL^2_1  \Phi_{\alpha_1}\bL_{-1}\bW_{-1}   |h_0, q_0 \rangle &=   6 \left(2 h_0+\left(h_1-1\right) h_1+1\right) q_0 \\& -\frac{1}{2} \left(h_1-1\right) \left(4 h_0+\left(h_1-1\right) h_1+2\right) q_1, \end{split} \\
& \langle  h_0, q_0|\bL^2_1\Phi_{\alpha_1} \bW_{-2}   |h_0, q_0 \rangle =   12 q_0-2 \left(h_1-1\right) h_1 q_1.
\end{split}
\end{equation}
\begin{equation}
\begin{split}
& \begin{split} \langle  h_0, q_0|\bW^2_1 \Phi_{\alpha_1} \bL^2_{-1}   |h_0, q_0 \rangle & =   18 q_0^2+6 \left(h_1-1\right) q_1 q_0+\frac{\left(h_1^3-7 h_1-30\right) q_1^2}{4 h_1} \\&-\frac{1}{5} h_1 \left(h_1 \left(h_1+3\right)-2\right)-\frac{6 h_0}{5},  \end{split} \\ 
&\begin{split} \langle  h_0, q_0| & \bW^2_1   \Phi_{\alpha_1}\bW^2_{-1}   |h_0, q_0 \rangle  =    \frac{\left(h_1-6\right) \left(h_1-3\right) \left(h_1+3\right) q_1^4}{16 h_1^3}\\&+\frac{\left(2 h_0 \left(h_1-3\right)-3 \left(h_1-2\right) h_1+12\right) q_1^2}{10 h_1}+\frac{1}{25} \left(2 h_0^2+\left(1-4 h_1\right) h_0+h_1 \left(3 h_1+2\right)\right) , \end{split}  \\
&  \begin{split} \langle  h_0, q_0| & \bW^2_1  \Phi_{\alpha_1}\bL_{-1}\bW_{-1}   |h_0, q_0 \rangle =   \frac{12}{40}  \left(\frac{15 q_1^2}{h_1}-5 q_1^2-4 h_0+4 h_1-6\right) q_0 +  \\&+  \frac{1}{40} \left(12 h_1^2-8 \left(h_0+1\right) h_1+8 h_0-\frac{5 \left(\left(h_1-3\right) \left(h_1-1\right) h_1-36\right) q_1^2}{h_1^2}+4\right) q_1,  \end{split} \\
& \langle  h_0, q_0|\bW^2_1\Phi_{\alpha_1} \bW_{-2}   |h_0, q_0 \rangle =   -\frac{\left(h_1-6\right) \left(h_1-3\right) q_1^3}{2 h_1^2}+\frac{2}{5} \left(2 h_1-3\right) q_1+\frac{6 q_0}{5}.
\end{split}
\end{equation}
\begin{equation}
\begin{split}
& \begin{split} \langle  h_0, q_0|\bW_1\bL_1 \Phi_{\alpha_1} \bL^2_{-1}   |h_0, q_0 \rangle & =  6 \left(2 h_0+\left(h_1-1\right) h_1+1\right) q_0+ \\& +\frac{1}{2} \left(h_1-1\right) \left(4 h_0+\left(h_1-1\right) h_1+2\right) q_1 ,   \end{split} \\ 
&\begin{split} \langle  h_0, q_0| & \bW_1\bL_1  \Phi_{\alpha_1} \bW^2_{-1}   |h_0, q_0 \rangle  =    \frac{12}{40}  \left(\frac{15 q_1^2}{h_1}-5 q_1^2-4 h_0+4 h_1-6\right) q_0+ \\&+      \frac{1}{40}\frac{\left(h_1-1\right) q_1 \left(5 \left(h_1-6\right) \left(h_1+3\right) q_1^2-4 h_1^2 \left(-2 h_0+3 h_1+2\right)\right)}{h_1^2}  , \end{split}  \\
&  \begin{split} \langle  h_0, q_0| & \bW_1\bL_1 \Phi_{\alpha_1}\bL_{-1}\bW_{-1}   |h_0, q_0 \rangle =    \frac{1}{5} \left(-2 h_0^2-\left(h_1-2\right) \left(h_1-1\right) h_0+45 q_0^2+h_1 \left(\left(h_1-1\right) h_1+2\right)\right) \\&    -\frac{\left(h_1-3\right) \left(2 h_0+\left(h_1-1\right) h_1+2\right) q_1^2}{4 h_1} ,  \end{split} \\
& \langle  h_0, q_0|\bW_1\bL_1\Phi_{\alpha_1} \bW_{-2}   |h_0, q_0 \rangle =   \frac{2}{5} \left(h_1^2+h_1-2 h_0\right)-\frac{\left(\left(h_1-6\right) h_1+3\right) q_1^2}{h_1}. 
\end{split}
\end{equation}
\begin{equation}
\begin{split}
& \langle  h_0, q_0|\bW_2 \Phi_{\alpha_1} \bL^2_{-1}   |h_0, q_0 \rangle=   2 \left(6 q_0+\left(h_1-1\right) h_1 q_1\right) ,\\ 
&\begin{split} \langle  h_0, q_0|\bW_2 \Phi_{\alpha_1} \bW^2_{-1}   |h_0, q_0 \rangle & =   \frac{\left(h_1^2-9\right) q_1^3}{2 h_1^2}-\frac{3}{5} \left(h_1-1\right) q_1+\frac{6 q_0}{5}, \end{split}   \\
&  \begin{split} \langle  h_0, q_0|\bW_2  \Phi_{\alpha_1}\bL_{-1}\bW_{-1}   |h_0, q_0 \rangle &=   \frac{1}{10} \left(\frac{5 \left(h_1 \left(3-2 h_1\right)+3\right) q_1^2}{h_1}-8 h_0+2 h_1 \left(h_1+3\right)\right), \end{split} \\
& \langle  h_0, q_0|\bW_2\Phi_{\alpha_1} \bW_{-2}   |h_0, q_0 \rangle =    \frac{8 h_0}{5}-\frac{4 \left(h_1-3\right) q_1^2}{h_1} .
\end{split}
\end{equation}
\section{Weyl character formula}  
\label{apweylchafor} 
Here, we want to compute the character $\chi^{\mathfrak{sl}_3}_j$~\eqref{chasl3} of a finite-dimensional representation of $\mathfrak{sl}_3$ with highest-weight vector given by $j= m_1 w_1 + m_2 w_2$, where $w_{1,2}$  are the fundamental weights and $m_1, m_2$ are positive integers. Let us define some notations. The fundamental weights satisfy the following products 
\begin{equation}  
\begin{split}
& (w_1, w_1)=  (w_2, w_2) = \frac{2}{3}, \qquad  (w_1, w_2) = \frac{1}{3} \; ,
\end{split}
\label{wchafor1} 
\end{equation} 
The Weyl vector is defined as
\begin{equation} 
\rho = w_1 +w_2 \; ,
\label{wchafor2} 
\end{equation}
and the Weyl group $W$ of $\mathfrak{sl}_3$ is formed by 6 elements, which are denoted as
\begin{equation}
W=[\Id,\; s_1,\; s_2,\; s_1s_2,\; s_2s_1,\; s_1s_2s_1] \; .
\label{wchafor3} 
\end{equation}
and they act on the vector $j$ as follows
\begin{equation} 
\begin{split}
\Id(j)	& = j \; ,\\ 
s_1(j)		& =  -m_1 w_1+ (m_1+m_2)w_2 \; ,\\ 
s_2(j)		& =   (m_1+m_2)w_1-m_2 w_2 \; ,\\ 
s_1s_2( j)		& =  - (m_1+m_2) w_1+ m_1 w_2 \; ,\\
s_2s_1 (j)		& =  m_2 w_1-(m_1+m_2) w_2 \; ,\\
s_1s_2s_1( j)	& = - m_1 w_1- m_2 w_2 \; .
\end{split}
\label{wchafor4} 
\end{equation}
The signature $\epsilon(w)$ of a element $w \in W$ is defined by
\begin{equation}  
\epsilon(w) = (-1)^{l(w)}\; , 
\label{prisu05} 
\end{equation} 
where $l(w)$ is the length of $w$, that is the number of $s_i$ that $w$ contains. For example, the signatures of the elements~\eqref{wchafor3} are respectively
\begin{equation}  
\epsilon(W)= [1,\; -1,\; -1,\; 1,\; 1,\; -1] \; .
\label{prisu06} 
\end{equation}
We compute the character $\chi^{\mathfrak{sl}_3}_j$ from the Weyl character formula
\begin{equation} \label{theprigene04}
\chi^{\mathfrak{sl}_3}_j = \frac{\sum_{w\in W} \epsilon(w)e^{(w(j+\rho), 2\pi i \tau \rho)}}{\sum_{w\in W} \epsilon(w)e^{(w(\rho),2\pi i \tau \rho) }} \; ,
\end{equation}
where we use $q= e^{2 \pi i \tau}$, obtaining exactly~\eqref{chasl3}.

\section{AGT relation} 
\label{apagtrelation} 
In this section, we comment on some important details of the AGT relation we used to compute the expression~\eqref{w3cb}. According to the AGT relation, the $\mathcal{W}_3$ conformal block $\mathcal{F}(\alpha, \alpha_1, q)$~\eqref{w3cb} can be computed by
\begin{equation} 
\mathcal{F}(\alpha, \alpha_1, q)= Z^{\inst}_{SU(3)}  \prod^{\infty}_{i=1}  (1-q^i)^{1 -2h_1 } \; ,
\label{apagt1} 
\end{equation}
where the infinite product is the so-called \emph{Heisenberg factor}, and $Z^{\inst}_{SU(3)}$ is the $SU(3)$ instanton partition function. For the algorithm which we used to compute $Z^{\inst}_{SU(3)}$ we refer to~\cite{Belavin:2015ria}, specifically the section 3.2.1.  Here, we want to clarify the relations between the parameters of $\mathcal{F}(\alpha, \alpha_1, q)$  and  $Z^{\inst}_{SU(3)}$. 

The partition function $Z^{\inst}_{SU(3)}$ is given by equation (28) of~\cite{Belavin:2015ria} and depends on the parameters $x_i= (Q  - \alpha, e_i), \mu, \epsilon_1, \epsilon_2$ which are related to our parameters as follows
\begin{equation}
\mu= a/3, \quad \epsilon_1= b, \quad \epsilon_2=1/b,
\end{equation}
while $\alpha$ is given by (\ref{alpha2}) and $Q, e_i$ are according to our notations above. Finally, the expression~\eqref{w3cb} is given in the limit $ c \rightarrow \infty$ (or equivalently $b \rightarrow 0$) of~\ref{apagt1}.

\section{Examples} 
\label{wilexamples} 
The simplest test\footnote{Since the expression (\ref{1pt2}) contains contributions of two states to the first level and four states to the second level, then $m_1, m_2 \ge 2$.} of~\eqref{mainequation} we can perform is the case $j= 2 w_1  + 2 w_2 = (2,2), j_1 = (3,0)$. In this case the tensor~\eqref{thesymmetrictensorM} becomes
\begin{equation}
^{b_1 b_{2}}_{a_1  a_{2}}M^{c_1  c_{2}}_{d_1  d_{2}}  =  \frac{1}{4} \bar{p}^{b_2} (D_{a_1 d_1}p_{d2}+ D_{a_1 d_2}p_{d1}) (\bar{D}^{b_1 c_1} A_{a_2}^{c_2} + \bar{D}^{b_1 c_2} A_{a_2}^{c_1} ) \; . 
\end{equation}
The next step is to make this tensor traceless according to~\eqref{eqn:OCB5} and then take the contraction according to~\eqref{mainequation}. We can show that there will be 16 terms proportional to $\mathcal{C}_1$, and all the terms proportional to $\mathcal{C}_2$ vanish. To write the result, we use the notation of the matrix $B$
\begin{equation}
B^{a}_{b}=\bar{p}^{a}p_b \; . 
\end{equation}
By using this notation, and denoting the trace of a generic matrix $C$ by $\Tr[C]= [C]$, one can show that
\begin{align}
&  {}_{a_1a_{2}}^{b_1  b_{2}}\tilde{M}_{(d_1 d_{2})}^{(c_1  c_{2})}\delta_{b_1 c_1}  \delta_{b_{2} c_{2}}\delta^{a_1 d_1}  \delta^{a_2 d_2} = \displaybreak[0] \notag \\
& = \frac{1}{4}\bigg( \mathcal{C}_0 \Big{(} [D]^2 [A B]+[D] [\bar{D}B A] + [\bar{D}] [DA B] + [D A \bar{D}B]  \Big{)}  +\frac{\mathcal{C}_1 }{4} \Big{(} [D \bar{D} B A^{T}]   +[D AB \bar{D}] \displaybreak[0] \notag \\ 
&\quad + [D \bar{D} B A] + [\bar{D}D][AB] +[D][\bar{D}B A^{T}] +[D][B \bar{D}A] +[\bar{D}B  DA^{T}] + [D \bar{D}B AB]\displaybreak[0] \notag \\
&\quad + [\bar{D}][DA B^{T}]+[\bar{D}][DA^{T}B]+[D A \bar{D} B^{T}]+[D \bar{D}B A^{T}B^{T}]+ [\bar{D}][D][A^T B] \displaybreak[0] \notag \\
&\quad + [D][\bar{D} A^{T}B]+ [\bar{D}][B D A^{T}]+[D \bar{D}A^T B] \Big{)}   +(0)\mathcal{C}_2  \bigg{)} = \displaybreak[0] \notag \\
& = \tilde{C}(2,2,3)(1-\frac{8}{5}q^2- \frac{4}{5}q^3 + \frac{4}{5}q^5 + \frac{8}{5}q^6 -q^8)\; .
\end{align}
The firs three  terms in $q$, (i.e. up to $q^2$) of the above expression are exactly the terms obtained by~\eqref{1pt2} in this case ($m_1, m_2, a\rightarrow  (2,2,3)$). This validates~\eqref{mainequation}. 

It is clear that the difficulty of this ``simple'' problem increases substantially when $m_1$, $m_2$, $a$ increase. We were able to work out the cases~\eqref{ourexamples}. Here we give the results and let us emphasize that all these cases confirm~\eqref{mainequation}.
\paragraph{Cases : } 
\begin{description}
\item[$(m_1,m_2, a)= (2,2,6)$] 
\begin{equation}
 {}_{a_1a_{2}}^{b_1  b_{2}}\tilde{M}_{(a_1 a_{2})}^{(b_1  b_{2})}  = \tilde{C}(2,2,6) (1-4q+4q^2+ \dots) \; ,
\end{equation}
\item[$(m_1,m_2, a)= (2,3,3)$]
\begin{equation}
 {}_{a_1a_{2}}^{b_1  b_{2}b_3}\tilde{M}_{(a_1 a_{2})}^{(b_1  b_{2}b_3)}  = \tilde{C}(2,3,3) (1+\frac{q}{3}-\frac{7 q^2}{9}+ \dots) \; ,
\end{equation}
\item[$(m_1,m_2, a)= (2,3,6)$]
\begin{equation}
 {}_{a_1a_{2}}^{b_1  b_{2}b_3}\tilde{M}_{(a_1 a_{2})}^{(b_1  b_{2}b_3)}  = \tilde{C}(2,3,6) (1-3q+q^2+\dots) \; ,
\end{equation}
\item[$(m_1,m_2, a)= (3,3,3)$]
\begin{equation}
 {}_{a_1a_{2}a_3}^{b_1  b_{2}b_3}\tilde{M}_{(a_1 a_{2}a_3)}^{(b_1  b_{2}b_3)}  = \tilde{C}(3,3,3) (1+\frac{2q}{3}+\frac{2q^2}{21}+\dots)\; ,
\end{equation}
\item[$(m_1,m_2, a)=(3,3,6)$]
\begin{equation}
 {}_{a_1a_{2}a_3}^{b_1  b_{2}b_3}\tilde{M}_{(a_1 a_{2}a_3)}^{(b_1  b_{2}b_3)}  = \tilde{C}(3,3,6) (1-2q-\frac{10q^2}{7}+ \dots) \, , 
\end{equation}
\item[$(m_1,m_2, a)= (4,4,3)$]
\begin{equation}
 {}_{a_1a_{2}a_3a_4}^{b_1  b_{2}b_3b_4}\tilde{M}_{(a_1 a_{2}a_3d_4)}^{(b_1  b_{2}b_3b_4)}  = \tilde{C}(4,4,3) (1+q+q^2+ \dots) \; , 
\end{equation}
\item[$(m_1,m_2, a)= (4,4,6)$]
\begin{equation}
 {}_{a_1a_{2}a_3a_4}^{b_1  b_{2}b_3b_4}\tilde{M}_{(a_1 a_{2}a_3a_4)}^{(b_1  b_{2}b_3b_4)}  = \tilde{C}(4,4,6) (1-q-\frac{5q^2}{3}+ \dots) \; .
\end{equation}
\end{description}

\bibliographystyle{JHEP} 
\bibliography{refs} 
\end{document}